# DISCUSSING THE DISCREPANCY BETWEEN ANALYTICAL CALCULATIONS AND THE OBSERVED BIOLOGICAL EFFECTIVENESS IN PROTON BORON CAPTURE THERAPY (PBCT)

**G.A.P. Cirrone[1,2], G. Petringa[1, *], A. Attili[3], D. Chiappara[1,6], L.Manti[4,5], V. Bravatà[7], D. Margarone[2], M.Mazzocco[6,8] and G. Cuttone[1]**

[1]Laboratori Nazionali del Sud – Istituto Nazionale di Fisica Nucleare, Via S Sofia 63, 95123, Catania (I)

[2]Institute of Physics Czech Academy of Science ELI-Beamlines Za Radnicí 835 Dolní Břežany (CZ)

[3]INFN Section of Roma "Roma Tre", Rome (I)

[4]Physics Department, University of Naples Federico II, Naples, Italy

[5]INFN Naples Section, Complesso Universitario di Monte S. Angelo, Via Cintia, Naples, Italy

[6]Dipartimento di Fisica e Astronomia, Università di Padova, via F. Marzolo 8, I-35131 Padova, Italy

Institute of Molecular Bioimaging and Physiology - National Research Council - (IBFM-CNR), Cefalù, Italy

[8]INFN-Sezione di Padova, via F. Marzolo 8, I-35131 Padova, Italy

*Abstract. A work recently published experimentally demonstrates an increase in the radiobiological efficacy of clinical proton beams when a tumour is treated in the presence of a concentration of [11]B. The paper, for the first time, demonstrates the potential role of the p+[11]B —> 3α (for brevity, p-B) reaction in the biological enhancement of proton therapy effectiveness. The work reports robust experimental data in terms of clonogenic cell survival and chromosomal aberrations and unambiguously shows the presence of an enhancement when cells were exposed to a clinical proton beam subject to treatment with sodium boroncaptate (BSH). Moreover, the greater occurrence of complex-type chromosomal exchanges points to the effect as due radiation of a LET (Linear Energy Transfer) greater than that of protons alone, possibly the alpha particles generated by the reaction. At the same time, we emphasized that analytical calculations, performed on the basis of the well-known total production cross section data, are not able to explain the effect in a macroscopic way, i.e. solely in terms of a trivial increase in the total dose released in the cells by the alpha-particles. In this paper, thanks to simulations and analytical calculations, we will discuss the theoretically expected alpha particle yield and the corresponding LET and RBE (Relative Biological Effectiveness) increase related to the [11]B presence. We conclude that a mere calculation based on the classical concepts of integral Dose, and average LET and RBE cannot be used to justify the observed radiobiological phenomena. We therefore suggest that micro- and nano-dosimetric aspects must be taken into account.*

---

*

*Corresponding author: giada.petringa@lns.infn.it

1. INTRODUCTION

Protontherapy represents one of the fastest growing radiotherapy modalities and one of the most attractive solutions when radiotherapy is needed, even compared to newer and advanced photon-based techniques such as intensity modulated radiation therapy (IMRT).

Protons are considered a low-LET (Linear Energy Transfer) radiation, with a clinical Relative Biological Effectiveness (RBE), assumed to be 1.1 (only ten percent bigger than gamma or x-ray radiation). Such a low biological efficiency is considered a limit especially in all those cases where radioresistant tumors must be treated, for which higher Z ions are preferred (i.e $^{12}$C ions). For these reasons, the protontherapy community is actively investigating the possibility to increase the proton biological efficiency by the concomitant use of different drugs and/or other agents to radiosensitize the tumor.

In this paper, the possibility to take advantage of $^{11}$B isotope as an agent able to deliver secondary particles enhancing the radiobiological effectiveness of protons triggering nuclear reactions, is discussed.

The clinical proton-Boron (p-B) approach was first proposed by Do-Kun et al. [1]. They suggested to exploit the features of the p($^{11}$B,α)2α fusion reaction, where low-energy (about 4 MeV) alpha particles are generated with a not negligible production cross section (around 1.7 barn) for low-energy (resonance at about 650 keV) incident protons.

The Korean group also discussed the possibility to use a 719 keV prompt-gamma, emitted in the interaction of proton with $^{11}$B as an imaging source and potential indicator of $^{11}$B uptake region. Giuffrida et al. [2] and Petringa et al. [3], on the opposite, underlined that the gamma emission is not generated from the fusion process but from the elastic p($^{10}$B, $^{10}$B')p' reaction (when using a mixture of $^{10}$B and $^{11}$B) and from the p($^{11}$B, $^{11}$B')p' channel and that, in the latter case, at the clinical attainable $^{11}$B concentration in tissues (order of 10$^{-5}$ g/cm$^3$), the peak signal would be completely masked by the non-prompt components generated as protons interact with human body natural components ($^{16}$O mainly).

The first experimental proof of radiobiological enhancement obtained by using the p-B reaction has been published by Cirrone et al. [4] irradiating prostate cancer cells (DU145) and non-tumorigenic breast epithelial MCF-10A cell lines along a 62 MeV clinical proton spread out Bragg peak (SOBP) at the Italian ocular protontherapy facility of INFN-LNS, Catania, Italy. Both clonogenic dose-response curves and complex type chromosomal aberration analyses clearly showed that the presence of Boron nuclei results in an increase of the radiobiological effectiveness of the proton beam. Besides the advantage of using a neutron-free nuclear fusion reaction, the relevance of this method stems from the fact that the reaction cross-section becomes significantly high at relatively low incident proton energy, *i.e.* around the Bragg peak region. In conventional protontherapy, the beam is typically slowed down inside the tumor (the Bragg peak region). Thus, most of the beam energy (related to the total released dose) is delivered to the tumor cells. Assuming that a given concentration of $^{11}$B nuclei is present preferentially, but not exclusively, in the tumor volume, the incoming slow protons can trigger fusion reaction events and generate highly DNA-damaging alpha particles.

The observed radiobiological enhancement reported in [4], even if confirmed in four experimental sessions, cannot be still explained making simple use of the reaction cross sections knowledge and/or of analytical considerations based on the classical concepts of dose, LET and RBE.

In this paper, the calculation of alpha particles along the proton SOBP in the irradiation conditions used in [4], will be presented in detail and discussed.

The recent radiobiological experimental results on PBCT are briefly recalled in Section 2 while the estimation of the generated alpha particles along the SOBP and how these affect the LET distributions are reported in Section 3 and Section 4.

2. THE RADIOBIOLOGICAL EVIDENCE

In order to verify the potentiality of the $^{11}$B(p, α)2α reaction for the enhancement of cell killing in a therapeutic protontherapy scenario, cells from the human prostate cancer line DU145 cell line have been irradiated at graded doses (0.5, 1, 2, 4 Gy) in the middle of a 62-MeV clinical SOBP.

A significant effect due to the Boron treatment (dissolved in the medium 6-8 hr prior to exposure as to give a concentration of 80 ppm) was observed throughout four experimental campaigns [4]. Based upon the measured clonogenic survival dose-responses, a Dose Modification Factor (or DMF) of 1.46 ± 0.12 was determined at the 10% survival level. This has suggested that the Boron atoms could confer a radiobiological advantage in cell killing

Investigation of structural DNA damages in the form of chromosome aberrations (CAs), a well-acknowledged biomarker of high-LET radiation exposure, subsequently confirmed the presence of high-LET radiation strongly advocating their contribution to the clonogenic results.

3. IMAGING POTENTIALITIES OF P-$^{11}$B REACTION

The presence of Boron within or in the vicinity of the neoplastic cells could also be exploited to spatially map the biological distribution of Boron nuclei and, hence, the tumor regions where the RBE enhancement occurs [1]. In particular, the gamma emitted in the proton elastic scattering with an $^{11}$B nucleus, generates a 718 keV prompt-gamma. The possibility to detect such gamma emission could allow $^{11}$B localization and, consequently, identification of the proton beam interaction site.

Simulations and experimental results conducted by Petringa et al. [3] demonstrated that, if a realistic $^{11}$B concentration is assumed, such prompt-gamma emission is impossible to be detected, being completely covered by the background gamma rays produced by proton interaction with other constituents of the human body.

In order to maintain proton-Boron therapeutic efficiency and, simultaneously, its imaging capability, the use of Boron coupled with natural Copper was proposed [5]. Copper can be easily bound to the dipyrromethene, BodiPy, a carrier molecule used for biological labelling, drug delivery and imaging. Natural Copper consists of two stable isotopes ($^{63}Cu$ and $^{65}Cu$) that can emit several gamma prompt peaks in the energy range between 0 to 2 MeV, mostly generated in the $^{63}Cu(p,p')^{63}Cu$ elastic reactions. A combination of different prompt-gamma rays could be useful for a mapping of the dose. Specifically, the 1.3 MeV gamma line could be used, being energetically located very far from the spectral region where emission of gamma from biological components is present.

## 4. MONTE CARLO SIMULATIONS AND ANALYTICAL EVALUATIONS TO ESTIMATE THE ALPHA YIELD

Monte Carlo (MC) simulations coupled to analytical calculations were adopted to estimate the total alpha particles produced at the cell site of the experiment reported in Cirrone et al. [4] in the attempt to explain the observed radiobiological enhancement. The proton spectra at various depths in water have been reproduced using MC; these, coupled with the alpha total production cross section, allowed to derive analytically the expected alpha particle yield. The total production cross section was derived from experimental data (using the EXFOR [6] database) and TALYS [7] calculation, when experimental data were not available.

### 4.1. Monte Carlo simulations

Simulations were carried out using the Geant4 (GEometry ANd Tracking) toolkit [8], version 10.04.p02 and the official Geant4 advanced example Hadrontherapy [9]. This example is able to simulate each element of the CATANA (Centro di Adroterapia ed Applicazioni Nucleari Avanzate) passive proton beam line of the INFN-LNS, where the experimental campaigns reported in [4] were performed.

The simulated geometry includes all the transport elements comprising the specific energy modulator to generate the SOBP, diagnostic elements and the water T-25 flask where cells were positioned. The latter is simulated through a box, 40x40x40 mm in dimension, divided into slices of water, 100 μm in thickness. Slices are perpendicular to the beam axis direction. The number of alpha particles was computed in the slices along the whole SOBP.

All the necessary physics processes were included in the simulations: decay physics, electromagnetic physics and physics of the hadronic (elastic and non-elastic) interactions. For hadronic processes, the Geant4 QGSP_BIC physics list was used [10]. Electromagnetic physics was included activating the G4EmStandardPhysics_opt4 physics list, able to provide high accuracy in the electrons, hadrons and ions tracking [11].

A total of 1M proton histories have been tracked in each simulation run.

The simulated energy spectra of the primary proton beam are reported in Figure 1 at four different depths along the SOBP (0.0 mm, 19.0 mm, 25.0 mm, 29.4 mm in water).

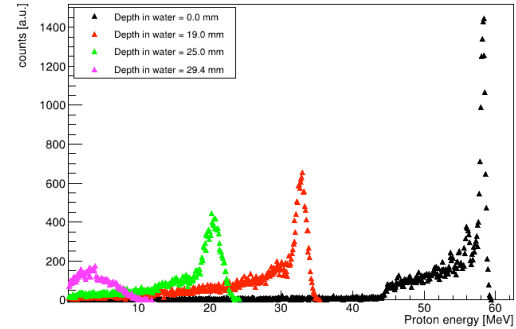

Figure 1. Protons energy spectra at various depths inside the water tank

### 4.2. Alpha production cross sections

Proton-Boron total production cross sections are needed for the computation of the expected number of produced alphas. TALYS-1.9 analytical code [7] was used to retrieve them due to its ability to account for all different production channels induced by protons (Figure 2).

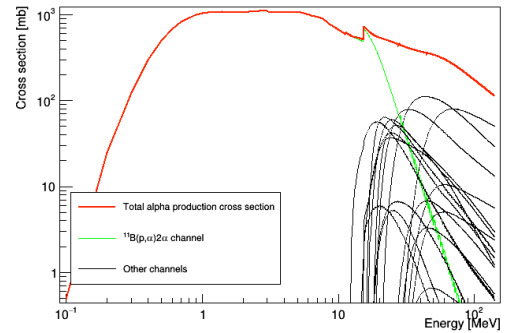

Figure 2. TALYS calculations for the total proton-Boron alpha production cross-section (red curve) plotted together with cross-section of other possible channels.

At low energies (below 10 MeV), the $^{11}B(p, \alpha)2\alpha$ channel (green curve of Figure 2) is the only one producing alpha particles; experimental cross sections for this channel are also available and data can be found in the EXFOR database [6]. Figure 3 shows the $^{11}B(p, \alpha)2\alpha$ total experimental cross section and the corresponding TALYS results when only this channel is considered.

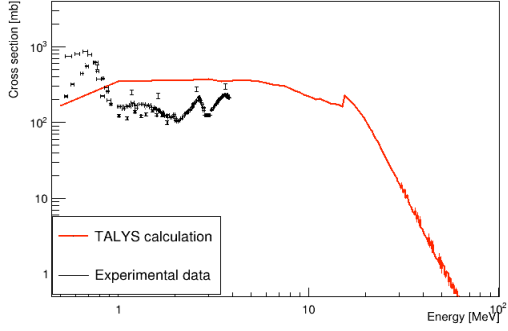

Figure 3. Experimental and TALYS $^{11}$B(p,a)2a total cross section vs energy of the incident proton

Considering that TALYS poorly describes the data below 1 MeV, we decided to adopt a combination of the two distributions as total cross section: experimental data below 4 MeV and TALYS calculations above that energy. The final resulting cross section is reported in Figure 4.

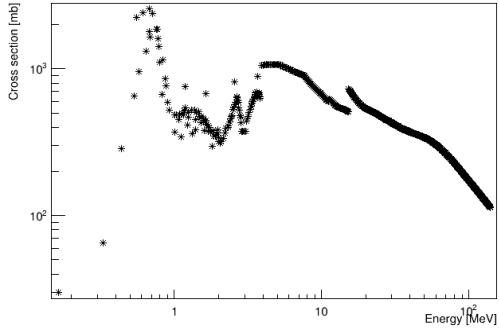

Figure 4. total alpha production cross section adopted for alphas calculation and derived by the combination of experimental and TALYS data

### 4.3. Alpha estimation

A quantification of alpha particles generated at any depths ($N_\alpha$) along the SOBP due to the $^{11}$B presence was performed embedding the simulated proton spectra of Figure 1, in the equation:

$$\frac{dN}{N} = \sigma_\alpha n_x d \quad (1)$$

where $\sigma_\alpha$ is the total alpha production cross section derived from the union of TALYS and experimental data (Figure 4), $n_x$ expressed in atoms/m$^3$ ($n_x$ = 3.56·10$^{24}$ atoms/m$^3$ corresponding to the $^{11}$B concentration used) and d is the voxel size (100 µm).

The $^{11}$B numerical density $n_x$ was estimated using the following equation:

$$n_x = f \frac{C_B \rho_{tot} N_A}{M_B} \quad (2)$$

where $C_B$ is the massic concentration of Boron (80 ppm in our case), $\rho_{tot}$ is the total density of the target (in this case it has been taken equal to water) $N_A$ is the Avogadro constant, $M_B$ is the atomic mass of Boron and f is the fraction of $^{11}$B with respect to the total natural Boron amount introduced into the cell culture through the BSH agent (here a fraction of 0.8 has been considered).

The final number of alpha particles is calculated by scaling the total number of proton $N_x$ in each slide in order to obtain a dose of 2 Gy in the middle of the SOBP.

In Figure 5 the estimated number of alphas, at different depths in water along the SOBP, is reported together with the experimental depth dose distribution.

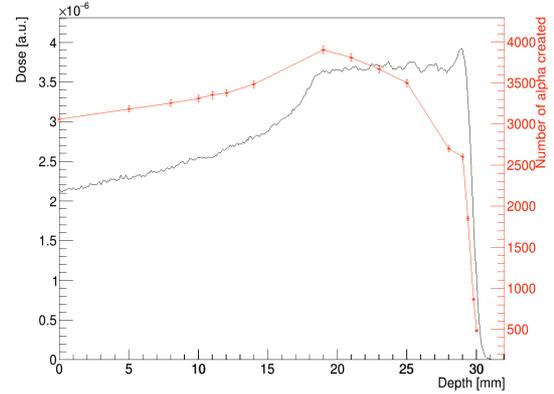

Figure 5: Number of alphas estimated as a function of depth in water inside the phantom (red curve) and the corresponding experimental SOBP. Lines connecting points are used as guideline.

### 5. DISCUSSION

The average Linear Energy Transfer (or LET) can be retrieved considering only the primary proton spectra along the SOBP (Figure 1) using the formula:

$$L_D = \frac{\sum L_i \varepsilon_i}{\sum \varepsilon_i} \quad (3)$$

here $L_i$ is the ratio between the deposited energy by the incident proton with an energy $\varepsilon_i$ and a step length $l_i$ [12].

The same definition of Equation 3 can be adopted to estimate the LET along the SOBP also including the contribution of the generated alphas $n_x$.

The calculated LET variation, after the inclusion of the generated alphas, appears, on the one hand, to be negligible and can be shown that, in order to obtain a DMF as the one experimentally observed, a LET of the order of 4-30 keV/µm should be observed. On the other hand, these LET values can be obtained only with a number of alpha particles larger by a factor 10$^2$ with respect to those reported in Figure 5. In Table 5 the proton LET, the total LET (including the generated alphas) and the corresponding DMF are reported for three different positions along the SOBP, also considering an additional alpha particle contribution which is a factor 10$^2$ larger with respect to the values reported in Figure 5. The calculated DMFs values are, in this case, closer to the experimental ones.

| Position | Depth in water [mm] | Proton LET [KeV/μm] | Total LET [KeV/μm] | Dose Modifying Factor (DMF) |
|---|---|---|---|---|
| Entrance | 5 | 1.96 | 3.66 | 1.17± 0.37 |
| Mid-SOBP | 21 | 4.24 | 9.53 | 1.36± 0.44 |
| Distal | 29.8 | 20.62 | 33.92 | 2.04 ± 0.66 |

Table 1: DMF estimation (last column) at three different depths along the SOBP; Proton dose averaged LET and Total dose averaged LET due to the contribution of alpha particles are also reported.

6. Conclusion

Inelastic reactions triggered by energetic protons as they slow down across a clinical SOBP interacting with Boron nuclei seem to produce an enhancement of proton radiobiological effectiveness. These results, even if experimentally well corroborated by repeated experimental campaigns [4], seem to fall short of an immediate justification by estimations merely based on analytical and Monte Carlo evaluations of classical dosimetric quantities such as integral dose, LET and RBE.

The total number of alpha particles generated, and estimated on the basis of the well-known total production cross section of the p-B reaction, does not explain the experimental results in terms of an average LET increase and, hence, with a corresponding RBE enhancement. On the opposite, if classical radiobiological models are applied, in order to reach the observed DMF values, one would require a number of alpha particles larger by a factor $10^3$ with respect to the calculated yield.

*Acknowledgement:* This paper has been written in the framework of the Move-it, MC-INFN and NEPTUNE projects funded by the Committee V of Italian Institute for Nuclear Physics (INFN).